\documentclass[a4paper,12pt]{article}
\usepackage{amssymb}
\usepackage{graphicx}

\def\journal#1#2#3#4#5{{\it #1} #2, #4 #3}

\def\v{\mathbf{v}}
\def\x{\mathbf{x}}
\def\u{\mathbf{U}}

\def\mb{\bar{m}}
\def\mt{\tilde{m}}
\def\lt{\tilde{\lambda}}
\def\xp{{x^\prime}}
\def\l{{\cal L}}
\def\cd{{\cal D}}

\def\z{{\cal Z}}

\def\cd{{\cal D}}
\def\cj{{\cal J}}
\def\pd{\partial}

\def\lt{\tilde{\lambda}}
\def\mt{\tilde{m}}
\def\at{\tilde{\alpha}}

\def\exp{\mathrm{exp}}

\def\be{\begin{equation}}
\def\ee{\end{equation}}
\def\bea{\begin{eqnarray}}
\def\eea{\end{eqnarray}}
\def\ie{\textit{i.e.} }

\title{The effects of bio-fluid on the internal motion of DNA}

\author{A. Sulaiman$^{a,b}$\footnote{Email : asulaiman@webmail.bppt.go.id, sulaiman@teori.fisika.lipi.go.id}
\, \, and \, \, 
L.T. Handoko$^{c,d}$\footnote{Email : handoko@teori.fisika.lipi.go.id, laksana.tri.handoko@lipi.go.id}}
\date{}

\begin{document}

\maketitle


\thispagestyle{empty}

\begin{center}
\begin{small}
\noindent
$^{a)}$Department of Physics, Bandung Institute of Technology\footnote{http://www.fi.itb.ac.id}, Jl. Ganesha 10, Bandung 40132, Indonesia\\
$^{b)}$P3 TISDA BPPT\footnote{http://tisda.bppt.go.id}, BPPT Bld. II (19$^{\rm th}$ floor), 
Jl. M.H. Thamrin 8, Jakarta 10340, Indonesia\\
$^{c)}$Group for Theoretical and Computational Physics, Research Center for Physics, Indonesian Institute of Sciences\footnote{http://teori.fisika.lipi.go.id}, Kompleks Puspiptek Serpong, Tangerang 15310, Indonesia\\
$^{d)}$Department of Physics, University of Indonesia\footnote{http://www.fisika.ui.ac.id}, Kampus UI Depok, Depok 16424, Indonesia\\
\end{small}
\end{center}

\vspace*{5mm}

\begin{abstract}
The internal motions of DNA immersed in bio-fluid are investigated. The interactions between the fragments of DNA and the surrounding bio-fluid are modeled using the gauge fluid lagrangian. In the model, the bio-fluid is coupled to the standard  gauge invariant bosonic lagrangian describing the DNA. It is shown that at non-relativistic limit various equation of motions, from the well-known Sine-Gordon equation to the simultaneous nonlinear equations, can be constructed within a single framework. The effects of bio-fluid are investigated for two cases : single and double stranded DNA. It is argued that the small and large amplitudes of a single stranded DNA motion immersed in bio-fluid can be explained in a natural way within the model as a solitonic wave regardless with the fluid velocity. In contrary the double stranded DNA behaves as regular or damped harmonic oscillator and is highly depending on the fluid velocity.
\end{abstract}

\vspace*{5mm}

\clearpage

\section{Introduction}

Both deoxyribo- and ribo-nucleic acid (DNA and RNA) have been recognized as the most important biomolecules. Especially DNA helical structures undergo a very complex dynamics which plays several important roles  in various biological phenomena such as storage of  information, inheritance (replication, etc) and the usage of genetic information (transcription, etc). The importance of biopolymers like DNA/RNA is motivated by established observations that the homologous recombination is preceded by  recognition and local pairing of intact double stranded DNA fragments, rather than involving known recombination proteins. Therefore, it should be attributed to direct DNA-DNA interactions whose physical origin has not been understood \cite{burgess,weiner}. Experimentally, the physical properties of DNA/RNA have been measured in many works, for example : the DNA single-molecule  \cite{smith,lavery,strick}, double stranded DNA forming bubbles \cite{altan}, the DNA/RNA nucleoside and nucleotides \cite{peon}, the structural transitions of DNA through torques measurements \cite{bryant}, the thermodynamic fluctuations of DNA in a reacting system \cite{magde}, the stretching DNA with a receding meniscus \cite{bensimon}, the electrical transport through single DNA molecules \cite{porath} and so forth.

From physical point of view, a biopolymer like DNA molecule is considered as a
system consisting of many interacting matters in a particular configuration of
space-time. Some models treats this kind of DNA dynamics as the phenomena of
nonlinear excitations like soliton. This type of models has been  pioneered by
Englander et.al. using nonlinear dynamics relevant to the transcription process
in terms of coupled pendulum chain  which generates the sine-Gordon equation and
its classical solitons \cite{englander}. Further, Davydov described the alpha
helices in quantum solitons \cite{davydov}. Following  these suggestions, a
number of models for the nonlinear DNA have been elaborated in the last decades,
in both classical or quantum approaches \cite{yakushevich, peyrard,cardoni}. A
typical classical approach is the so-called PDB model which takes into account
twisted DNA molecules \cite{peyrard2,dauxois,dauxois2}. On the other hand, there
are several models based on the particle interactions
\cite{lee,knotts,sulaiman}. Also, the polyelectrolyte model which treats DNA
molecule as a cylinder with a net charge homogeneously distributed along its
surface, and has further been modified to be the electrostatic zipper motif for
DNA aggregation \cite{kornyshev}, to solve high dependency of the electrostatic
interaction between DNA duplexer on surface charge patterns \cite{kornyshev2}.

It has also been shown that under particular external conditions the DNA molecules form a double helix, and its (transverse, longitudinal and torsional) motions can be divided into two main regions : the small and large amplitude of internal motions \cite{yakushevich2}. The small amplitude of motion can be described by the hamiltonian of harmonic oscillator. On the other hand, the large amplitude is described by a non-harmonic one \cite{mingalev}. Recently, many works have discussed and arrived at the conclusion that the large amplitude of internal motion can be considered as a nonlinear dynamical system where solitary conformational waves can be excited \cite{yakushevich}. Then nonlinear interaction between molecules in DNA gives rise to a very stable excitation as soliton \cite{mingalev,cardoni2}. 

As mentioned above, DNA is not motionless. It is in a constantly wriggling dynamics state in a medium of bio-organic fluid in the nucleus cell \cite{julia}. However, the motion of DNA surrounded by fluid is rarely studied. Previous studies are usually done by solving the fluid equations and its wave equations simultaneously using appropriate boundary conditions. On the other hand, in the Hamiltonian formulation the viscous force is considered to be comparable with other forces arising from Hamiltonian \cite{kovici,aslin}. The solution is then obtained by expansion and performing order-by-order calculation. In these approaches, anyway the picture of interaction between DNA and its surrounding fluid is not clear. Also, in most models the over-damped DNA dynamics are treated by putting some additional terms by hand in the differential equation to obtain the non-homogenous ones \cite{irwin}. The stochastic simulations of DNA in flow has been done for a fully parametrized bead–spring chain model by taking into account the fluctuating hydrodynamic interactions \cite{richard}.

In this paper, a new model to describe various internal motions of DNA inspired
by gauge fluid theory is proposed. The DNA dynamics is modeled as the result of
interactions  among matters in a fluid medium using the relativistic and gauge
invariant fluid  lagrangian. Although the theory is a relativistic one, we  take
its non-relativistic limit at the final stage to deal with problems in DNA as
done in some previous works, for instance in some models using the ideal gas
approximation \cite{fedyanin}. Moreover, the lagrangian is  intended for physics
at scale of order transport mean free paths, that is the transition region where
neither a hydrodynamics nor kinetic theory is valid. Therefore it fits the
current interest of modeling ``elementary'' biomatters like DNA. Just to
mention, the lagrangian is originally devoted to model the quark gluon plasma
(QGP) as a relativistic fluid system \cite{mahajan1,handoko2,mahajan2,handoko},
inspired by the similarity between the dynamical properties of fluid and
electromagnetic field \cite{marmanis,marmanis2}. The DNA is treated as strongly
coupled system like non-Abelian plasmas where neither a hydrodynamics nor
kinetic theory is really valid. Within the model, a single and double stranded
biopolymers are described in a general way as the results of interactions among
the fluid and matter fields. 

We show in two specific cases how to derive the equation of motion (EOM) and investigate the internal motions through its solutions and behaviors. From the EOM of a DNA as a single bulk, we argue that small and large amplitude regions of the internal motion of DNA are determined by its internal dynamics and interactions with surrounding fluid. On the other hand, in the case of double stranded DNA the EOM is solved analytically to investigate the effects of fluid velocity to its internal motion.

The paper is organized as follows. First we briefly introduce the theory of gauge invariant fluid lagrangian, and then provide the allowed interactions within the model. After explaining how to model DNA using the interactions in the lagrangian, we provide two typical examples : 1) the Abelian U($1$) case to model the dynamics of a single bulk of DNA, and 2) the non-Abelian SU($2$) case to describe the internal motion of double stranded DNA.  Finally, the paper is ended with summary and discussion.

\section{Theoretical background}

Here, a new approach to investigate the interaction between biopolymer and  its surrounding bio-fluid is discussed using the lagrangian method. Rather putting it by hand, the interaction is described in a more natural way from first principle, \ie by introducing some symmetries in the lagrangian under consideration.

\subsection{The lagrangian}

The model is an extension of the original model based on the U($1$) gauge theory
devoted for QGP as a magnetofluid system \cite{mahajan1,handoko2,mahajan2}.
Thereafter it has been extended to the non-Abelian case to accommodate a system
with many matters, either bosonic or fermionic ones \cite{handoko}. Concerning
the fact that an (elementary) matter has no intrinsic degree of freedom like
spin, it is considerable to represent its elementary constituents as  scalar
(boson) fields governed by the bosonic lagrangian, 
\be
        \l_\mathrm{matter} = \left( \pd_\mu \Phi \right)^\dagger  \left( \pd^\mu \Phi \right) + V(\Phi) \; , 
        \label{eq:lphi}
\ee
where $V(\Phi)$ is the potential. For example in the typical $\Phi^4-$theory, 
\be
	V(\Phi) = -\frac{1}{2} m_\Phi^2 \, \Phi^\dagger \Phi - \frac{1}{4!} \lambda \, (\Phi^\dagger \Phi)^2 \; ,
	\label{eq:v}
\ee
where $m_\Phi$ and $\lambda$ are the mass of matter and the dimensionless coupling constant of matter self-interaction. The hermite conjugate is  $\Phi^\dagger \equiv {(\Phi^\ast)}^T$ for a general complex field $\Phi$. 

We impose the above bosonic lagrangian to be gauge invariant under local (in general non-Abelian) gauge transformation \cite{yang,mills}, $U \equiv \exp[-i T^a \theta^a(x)] \approx 1 - i T^a \theta^a(x)$ with $\theta^a \ll 1$. $T^a$'s are generators belong to a particular Lie group and satisfy certain commutation  relation $[T^a,T^b] = i f^{abc} T^c$ with $f^{abc}$ is the anti-symmetric structure constant \cite{chengli}. The matter field is then transformed as $\Phi \stackrel{U}{\longrightarrow} \Phi^\prime \equiv \exp[-i T^a \theta^a(x)] \, \Phi$, with $T^a$ are $n \times n$ matrices while $\Phi$ is an $n \times 1$ multiplet containing $n$ elements, \ie 
\be
	\Phi = \left( 
	\begin{array}{c}
		\Phi_1 \\
		\Phi_2 \\
		\vdots \\
		\Phi_n \\
	\end{array}
	\right) 
	\; \; \; \mathrm{and} \; \; \; 
	\Phi^T = (\Phi_1 \; \Phi_2 \; \cdots \, \Phi_n) \; ,
	\label{eq:multiplet}
\ee
for $n$ dimension Lie groups as SU($n$), O($n+1$),  etc. It is well-known that the symmetry in Eq. (\ref{eq:lphi}) is revealed by introducing gauge fields $A_\mu^a$ which are transformed as $U^a_\mu \stackrel{U}{\longrightarrow} {U^a_\mu}^\prime \equiv U^a_\mu - \frac{1}{g}  (\pd_\mu \theta^a) + f^{abc} \theta^b U^c_\mu$, and replacing the derivative with the covariant one, $\cd_\mu \equiv \pd_\mu + i g \, T^a U^a_\mu$. Anyway, the number of generators, and also gauge bosons, is determined by the dimension of group under consideration. For an SU($n$) group one has $n^2 - 1$ generators and the index $a$ runs over $1, 2, \cdots, n^2 - 1$. For example the SU(2) group is realized by $2 \times 2$ matrices $T^a \equiv \frac{1}{2} \sigma^a$ with $\sigma^a$ are the Pauli matrices \cite{chengli},
\be
	\sigma^1 = \left(
		\begin{array}{cc}
		0 & 1 \\
		1 & 0 \\
		\end{array}
		\right) \; \; \; , \; \; \; 
	\sigma^2 = \left(
		\begin{array}{cc}
		0 & -i \\
		i & 0 \\
		\end{array}
		\right) \; \; \; , \; \; \; 
	\sigma^3 = \left(
		\begin{array}{cc}
		1 & 0 \\
		0 & -1 \\
		\end{array}
		\right) \; ,
	\label{eq:pm1}
\ee
In particular, the Abelian U($1$) case is revealed by putting $T^a \theta^a(x) \rightarrow \theta(x)$, \ie the phase transformation, respectively.

Finally, the gauge invariance leads to the total lagrangian with some additional terms in the lagrangian to keep its gauge invariance,
\be
	\l = \l_\mathrm{matter} + \l_\mathrm{gauge} + \l_\mathrm{int} \; ,
        \label{eq:l}
\ee
where, 
\bea
        \l_\mathrm{gauge} & = & -\frac{1}{4} S^a_{\mu\nu} {S^a}^{\mu\nu} \; ,
        \label{eq:la} \\
	\l_\mathrm{int} & = & -g J^a_\mu {U^a}^\mu + g^2 \left( \Phi^\dagger T^a T^b \Phi \right) U_\mu^a {U^b}^\mu \; .
        \label{eq:li}
\eea
The strength tensor is $S^a_{\mu\nu} \equiv \pd_\mu U^a_\nu - \pd_\nu U^a_\mu + g f^{abc} U^b_\mu U^c_\nu$, while the 4-vector current is,
\be
        J^a_\mu = -i \left[ (\pd_\mu \Phi)^\dagger T^a \Phi - \Phi^\dagger T^a (\pd_\mu \Phi) \right] \; .
        \label{eq:j}
\ee
The coupling constant $g$ then represents the interaction strength between gauge field and matter. We should note that,  the current conservation is realized by the covariant current $\pd^\mu \cj^a_\mu = 0$ with $\cj^a_\mu \equiv -i \left[ (\cd_\mu \Phi)^\dagger T^a \Phi - \Phi^\dagger T^a (\cd_\mu \Phi) \right]$ \cite{handoko}.

The gauge boson $U_\mu$ is  interpreted as a ``fluid field'' with velocity
$u_\mu$, and takes the form \cite{mahajan1,handoko2,mahajan2,handoko}, 
\be
 	U^a_\mu = \left( U_0^a, \u^a \right) \equiv u^a_\mu \, \phi \; ,
	\label{eq:a}
\ee
with, 
\be
        u_\mu \equiv \gamma^a (1, -\v^a) \; ,
        \label{eq:ae}
\ee
where $\phi$ is an auxiliary boson field, while $\gamma^a \equiv \left( 1 -
|\v^a|^2 \right)^{-1/2}$. Here we adopt the natural unit, \ie the light speed $c
= 1$. Eq. (\ref{eq:ae}) is nothing else than rewriting a gauge field in terms of
its polarization vector and wave function which represents the fluid
distribution in a system. It has further been shown that the non-relativistic
fluid equation can be reproduced using Eq. (\ref{eq:a}) \cite{handoko2,
handoko}. This fact actually justifies us to model the DNA dynamics in a fluid
medium using the total lagrangian in Eq. (\ref{eq:l}).

Now we are ready to model the DNA using the above lagrangian. First, we should investigate the allowed interactions in the present theory. 

\subsection{The interactions}

In order to be specific, let us consider the $\Phi^4-$potential in Eq. (\ref{eq:v}) for matter lagrangian in Eq. (\ref{eq:lphi}). With a complete lagrangian at hand, we can extract $m-$point interactions for fluid and matter with $m$ is the number of relevant legs involved in an interaction. We list all allowed interactions below for each element in the matter multiplet denoted by the indices $i,j$.
\begin{itemize}

\item $2-$point interactions :\\
The interactions arise through the kinetic and mass terms of matter in Eqs. (\ref{eq:lphi}) and (\ref{eq:v}), and the fluid kinetic term in Eq. (\ref{eq:la}),
\bea
	\Phi\Phi & : & \left( \partial_\mu \Phi_i^\ast \right)\left( \partial^\mu \Phi_i\right) 
		- \frac{1}{2} m_\Phi^2 \, \Phi^\ast_i \Phi_i \; .
	\label{eq:2p} \\
	UU 	& : & -\frac{1}{4} \left( \pd_\mu U^a_\nu - \pd_\nu U^a_\mu\right)\left( \pd^\mu {U^a}^\nu - \pd^\nu {U^a}^\mu\right) \; .
	\label{eq:2a} 
\eea

\item $3-$point interactions :\\
These interactions are induced by the fluid self-interaction in Eq. (\ref{eq:la}) and the fluid-matter interaction in Eq. (\ref{eq:li}), 
\bea
	\Phi\Phi U	& : & i g \, T^a_{ij} \left[ \left( \partial_\mu \Phi^\ast\right)_i \Phi_j - \Phi^\ast_i \left( \partial_\mu \Phi \right)_j \right]  {U^a}^\mu  \; .
	\label{eq:3pa} \\
	UUU       	& : & \frac{1}{2} g \, f^{abc} {U^b}^\mu {U^c}^\nu \left( \pd_\mu U^a_\nu - \pd_\nu U^a_\mu\right) \; , 
	\label{eq:3a}
\eea

\item $4-$point interactions :\\
These interactions are induced through the matter self-interaction in  Eq. (\ref{eq:lphi}), the fluid kinetic term in Eq. (\ref{eq:la}) and the fluid-matter interaction  in Eq. (\ref{eq:li}), 
\bea
	\Phi\Phi\Phi\Phi	& : & -\frac{1}{4!} \lambda \, \left( \Phi_i^\ast \Phi_i \right)^2 \; ,
	\label{eq:4p} \\
	UUUU			& : & -\frac{1}{4} g^2 \, f^{abc} f^{ade} {U^b}^\mu {U^c}^\nu U^d_\mu U^e_\nu \; , 
	\label{eq:4a} \\
	\Phi\Phi UU		& : & g^2 \,  \Phi^\ast_i \left( T^a T^b \right)_{ij} \Phi_j U_\mu^a {U^b}^\mu\; .
	\label{eq:4pa}
\eea

\end{itemize}

All of these constitute the so-called  Feynman diagrams and its order of magnitudes that will be used soon in the subsequent section. Now we are ready to construct the models relevant for biopolymers. 

\section{Modeling the DNA}

Here, we consider two typical examples on how to describe various dynamics of DNA within the present model. First we present a model for a single bulk of DNA or a fragment of DNA molecule like nucleotide or nucleoside. Further we construct a more complicated picture for the double stranded DNA. The model is a new type of the mesoscale model of DNA that reduces the complexity of a nucleotide to three interactions sites \cite{knotts}.

\subsection{Single bulk of DNA : the Abelian U($1$) model}
\label{subsec:u1}

The Abelian U($1$) lagrangian involves only a single matter and a fluid field.
In this case, the $3-$point interaction in Eq. (\ref{eq:3a}) and the $4-$point
interaction in Eq. (\ref{eq:4a}) vanish. It is also clear that we are not able
to construct a realistic model for a biopolymer composed by several different
matters in this case \cite{sulaiman}. However, we can model the dynamics of a
single bulk of DNA or its fragment like nucleoside which could be considered as
a composite field of sugar and base. This means we investigate the internal
dynamics of namely DNA molecules through the EOM of its fragments and study the
basic behaviors.

The total lagrangian in this case becomes, 
\bea
	\l & = & \left( \pd_\mu \Phi^\ast \right)  \left( \pd^\mu \Phi \right) - \frac{1}{2} m_\Phi^2 \, \Phi^\ast \Phi - \frac{1}{4!}\lambda \, \left(\Phi^\ast \Phi \right)^2 
	+ g^2 \, U_\mu U^\mu \, \Phi^\ast \Phi 
	\nonumber \\
	& & - \frac{1}{4} \left( \pd_\mu U_\nu - \pd_\nu U_\mu\right)\left( \pd^\mu {U}^\nu - \pd^\nu {U}^\mu\right) 
	+ i g \, U^\mu\ \left[ \left( \partial_\mu \Phi^\ast\right) \Phi - \Phi^\ast \left( \partial_\mu \Phi \right)\right] \; ,
	\label{eq:lu1}
\eea
using Eqs. (\ref{eq:v}) and (\ref{eq:l})$\sim$(\ref{eq:j}). Imposing the variational principle of action and the Euler-Lagrange equation in term of $\Phi$ \cite{chengli}, 
\be
	\frac{\partial \l}{\partial\Phi} - \partial_\mu \, \frac{\partial \l}{\partial\left( \partial_\mu \Phi \right) } = 0 \; ,
	\label{eq:ele}
\ee
we find the EOM for a single matter as follow, 
\be
	\left( \partial^2 + m_\Phi^2 + 2 g^2 \, U^2 \right) \Phi + \frac{1}{3!} \lambda \, \Phi^3 = 0 \; .
	\label{eq:eomu1}
\ee
for a real $\Phi$ field. 

This result leads to a solitonic wave equation for $\lambda \ne 0$ described by the well-known nonlinear Klein-Gordon equation, 
\be
	\left( \partial^2 + \mb_\Phi^2 \right) \Phi + \frac{1}{3!} \lambda \, \Phi^3 = 0 \; ,
	\label{eq:nlkge}
\ee
with $\mb_\Phi^2 \equiv m_\Phi^2 + 2 g^2 \, U^2$, and $U^2 = \phi^2$ from Eqs. (\ref{eq:a}) and (\ref{eq:ae}). Here $\lambda$ determines the 'level of non-linearity' for the Klein-Gordon equation. If one  puts $\lambda \approx \mb_\Phi^2$, we arrive at the sine-Gordon equation in $4-$dimensional space-time $(t,\x)$, $\partial_t^2 \Phi - \partial^2_\x\Phi - \mb_\Phi^2 \sin \Phi = 0$ using $\sin \Phi \approx \Phi - \frac{1}{3!} \Phi^3 + \cdots$. This kind of equation often appears in the models based on the coupled pendulum chains pioneered by Englander et.al.  \cite{englander}. However, we should note that the equality $\lambda \approx \mb_\Phi^2$ in this model doesn't make sense since $\lambda$ and $\mb_\Phi^2$ have different dimensions. In this paper, rather than considering that special case, let us solve Eq. (\ref{eq:nlkge}) in a general way.

For the sake of simplicity, we consider a traveling wave in $2-$dimensional space-time $(t,x)$, \ie $\Phi(\xp) \equiv  \Phi(x-Ct)$,  where $C$ is a phase velocity. Since $\partial^2_t\Phi = C^2 \partial^2_\xp \Phi$ and $\partial^2_x\Phi = \partial^2_\xp \Phi$, Eq. (\ref{eq:nlkge}) can be rewritten as,
\be
	\partial^2_\xp \Phi + \mt_\Phi^2 \, \Phi + \lt \, \Phi^3 = 0 \; ,
	\label{eq:nlkge2}
\ee
with $\mt_\Phi^2 \equiv {\mb_\Phi^2}/{(C^2 - 1)}$ and $\lt \equiv \lambda/{[3!(C^2 - 1)]}$. Assuming that $v_x = v$ is a constant makes $\mt_\Phi$ to also be a constant. Hence we can multiply both sides of Eq. (\ref{eq:nlkge2}) with $\partial_\xp \Phi$ to obtain,
\be
	\partial_\xp \left[ \left( \partial_\xp \Phi \right)^2 + \mt_\Phi^2 \, \Phi^2 + \frac{1}{2} \lt \, \Phi^4 \right] = 0 \; .
	\label{eq:nlkge3}
\ee
Concerning that the quantum wave function $\Phi$ has the Gaussian distribution, it is integrable and  then leads to the following differential equation,
\be
	\left( \partial_\xp \Phi \right)^2 + \mt_\Phi^2 \, \Phi^2 + \frac{1}{2} \lt \, \Phi^4 = 0 \; .
	\label{eq:nlkge4}
\ee
Through standard mathematical procedures, we can straightforwardly get the solution, 
\be
 	\Phi(x^\prime) = \left|\mt_\Phi \right| \sqrt{\frac{2}{\lt}} \, \mathrm{sech} \left( \left|\mt_\Phi \right| \, x^\prime \right) \; ,
	 \label{eq:soliton}
\ee 
for $\lt > 0$, or $|C| > 1$.

\begin{figure}[t]
        \centering 
	\includegraphics[width=11cm,angle=0,trim=0 0 0 0]{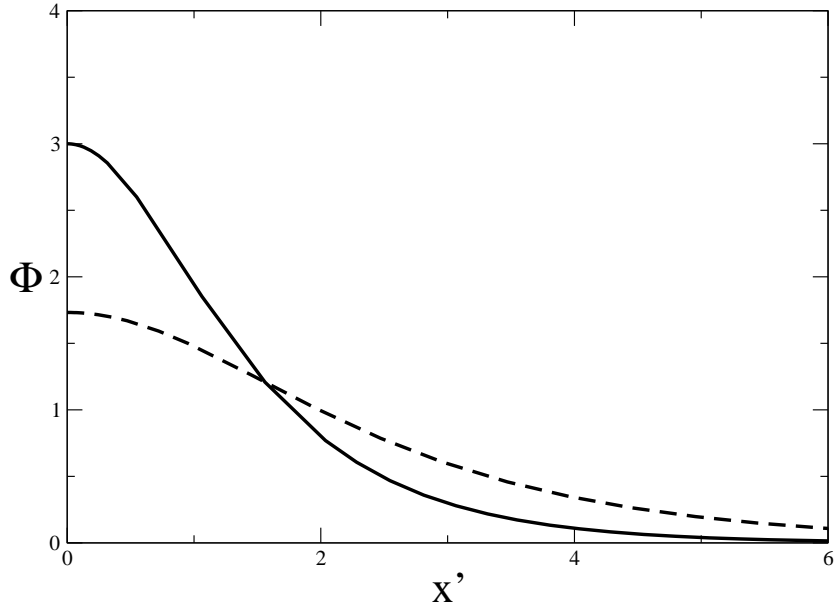}
        \caption{The solitonic wave function for a  2-dimensional DNA as a function of $x^\prime$ with the coupling constants $g = 0.1$ (solid line) and $g = 0$ (dashed line) for a fixed parameter set $(m_\Phi,\phi,C,\lambda) = (1,1,2,4)$.}
        \label{fig:soliton}
\end{figure}

The non-relativistic limit can be obtained by performing a transformation $t \rightarrow \tau \equiv i t$ in Eq. (\ref{eq:nlkge}), and  putting $\gamma \rightarrow 1$  respectively. This leads to the same result as Eq. (\ref{eq:soliton}), but $t$ is replaced with $-i \tau$. The behavior of this solitonic wave function is depicted in Fig. \ref{fig:soliton} as a function of $x^\prime$ with (solid line) and without (dashed line)  surrounding fluid for a fixed parameter set. Anyway, the fluid contribution is independent on its velocity $v$, since the effective mass $m_\Phi^2$ is shifted by $U^2 = \phi^2$. From the figure, we can conclude that the large and small amplitudes can be considered as the effects of fluid surrounding the DNA.

\subsection{Double stranded DNA : the non-Abelian SU($2$) model}
\label{subsec:su2}

Now let us apply the present lagrangian in a more realistic case of double stranded DNA. Concerning the smallest group beyond U($1$), we take the SU($2$) group to construct the model.  In this group, we have 2 sub-matters in a doublet of matter field as Eq. (\ref{eq:multiplet}) with $n = 2$.

\begin{figure}[t]
        \centering \includegraphics[width=12cm]{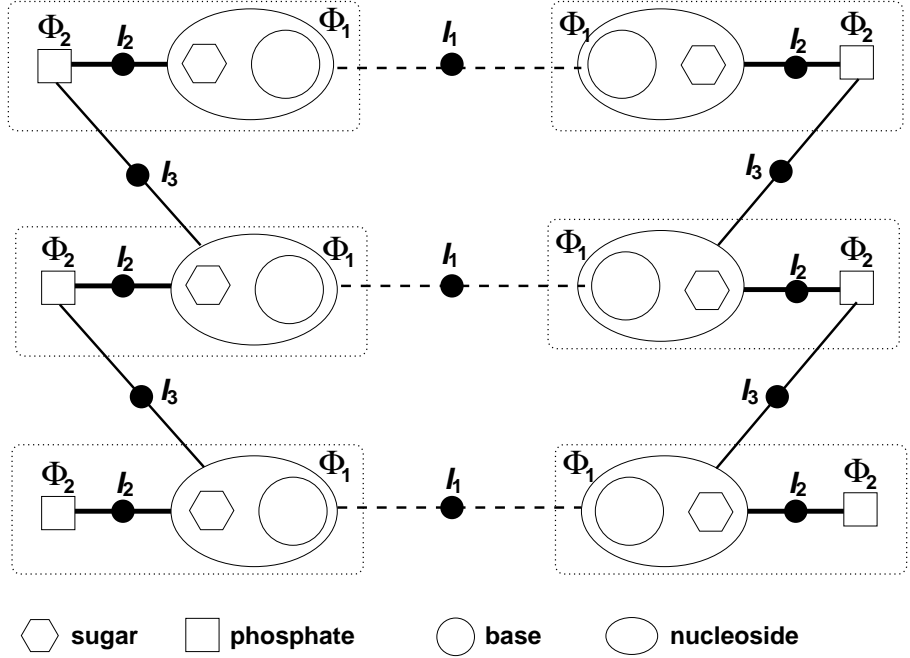}
        \caption{The double stranded DNA in the non-Abelian SU($2$) model with nucleosides and phosphates are represented by $\Phi_1$ and $\Phi_2$. The vertices  $I_1$, $I_2$ and $I_3$ denote different types of interactions connecting nucleosides ($\Phi_1-\Phi_1$) manifesting base pairs in neighboring strands, nucleoside$-$phosphate ($\Phi_1-\Phi_2$) within a nucleotide, and nucleoside$-$phosphate ($\Phi_1-\Phi_2$) between nucleotides in a strand.}
        \label{fig:model}
\end{figure}

\begin{figure}[t]
        \centering \includegraphics[width=15cm,trim=0 0 0 2cm]{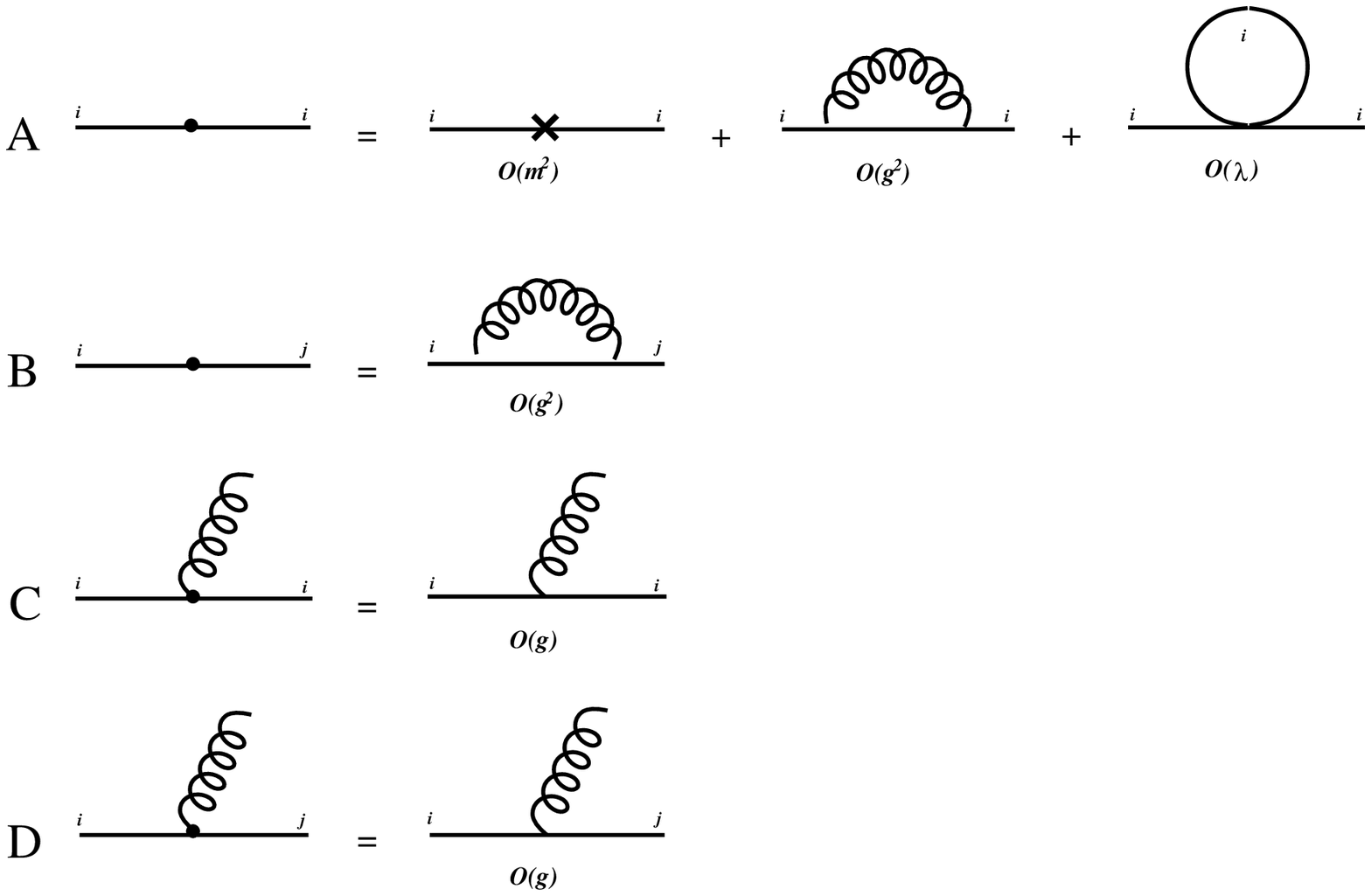}
        \caption{The $2-$point interactions and its first order contents relevant for double stranded DNA in Fig. \ref{fig:model} and its order of magnitudes. The plain and spring lines indicate matter and fluid fields.}
        \label{fig:fd}
\end{figure}

Since we have only 2 different states of matter, $\Phi_1$ and $\Phi_2$, it is convincing to split the nucleotide to be a phosphate and a nucleoside consisting of sugar and base. So, the interaction between two nucleotides, which further form the backbone of DNA molecule, is  attributed to the interaction of two different matters, \ie phosphate and nucleoside. On the other hand, the base pair is revealed as the interaction between two identical matters, \ie two neighboring nucleosides belonging to different strands. The model is schematically illustrated in Fig. \ref{fig:model} where we have assigned $\Phi_1$ for the nucleosides and $\Phi_2$ for the phosphates. Following the allowed interactions in Eqs. (\ref{eq:2p})$\sim$(\ref{eq:4pa}), we can easily estimate the order of magnitudes for each interaction relevant to Fig. \ref{fig:model} as listed in Fig. \ref{fig:fd}.

From Fig. \ref{fig:fd}, it is straightforward to deduce that $I_1$ bound is materialized by  vertex $A$, while vertex $B$ is responsible for $I_2$ and $I_3$ bounds. Anyway, we should note that there are another diagrams with spring loops in the vertices $A$ and $B$, however they  would be vanishing due to the anti-symmetric $f^{abc}$. Now, we unfortunately face a problem on distinguishing $I_2$ with $I_3$ in Fig. \ref{fig:model}. It is quite natural to consider $I_3$ must be larger than $I_2$, since the backbone should be rather strongly tied and rigid than the nucleotide. Therefore in order to resolve this problem we propose an additional contribution to $I_3$ coming from interacting fluid (either fluid absorption or emission) with  matters depicted in vertex $D$ of Fig. \ref{fig:fd}. Of course, so $I_1$ could receive additional contribution from vertex $C$ too. This scenario  could be understood in the following way. Since the backbone is more open to surrounding fluid  than the phosphate$-$nucleoside encaged in the nucleotide bound-state, its interaction with surrounding fluid would contributes more significantly, and then should be taken into account. 

We might remark that in the present case the nucleoside, consisting of sugar and base, should be  considered as a well-confined bound-state. So we are not going into insight to investigate its structure. In consequence of this, we can not distinguish the A$-$T (adenine$-$thymine) with the G$-$C (guanine$-$cytosine) base pairs. Although in principle, these might be explained using multi-loops gauge boson exchanges inside nucleosides, and two different base pairs are attributed to the fluid velocities in the fluid loops (the second diagram of vertex $A$ in Fig. \ref{fig:fd}) with opposite sign. However we postpone this point in this paper since it would require larger group like SU($3$) containing more matter fields. Anyway, the opposite torsional motions of neighboring strands forming a DNA molecule can be explained, at time being,  qualitatively by considering the surrounding fluid in both strands have the same velocities ($\v$) but with opposite sign each other. 

Now, we investigate the EOM in SU($2$) as done in Sec. \ref{subsec:u1}. Substituting the full lagrangian, Eqs. (\ref{eq:l})$\sim$(\ref{eq:li}), into Eq. (\ref{eq:ele}), we obtain for each element of matter,
\bea
	\partial^2 \Phi_1 + m_\Phi^2 \, \Phi_1 - 2 g \, \left( \partial_\mu U_2^\mu\right) \Phi_2 + \frac{1}{3!} \lambda \, \left( \Phi_1^2 + \Phi_2^2 \right) \Phi_1 - 4 g \, U_2^\mu \left( \partial_\mu\Phi_2 \right) & = & 0 \; ,
	\label{eq:eom1} \\
	\partial^2 \Phi_2 + m_\Phi^2 \, \Phi_2 + 2 g \, \left( \partial_\mu U_2^\mu\right) \Phi_1 + \frac{1}{3!} \lambda \, \left( \Phi_1^2 + \Phi_2^2 \right) \Phi_2 + 4 g \, U_2^\mu \left( \partial_\mu\Phi_1 \right) & = & 0 \; ,
	\label{eq:eom2}
\eea
for real fields $\Phi_i$ ($i : 1,2$). Eqs. (\ref{eq:pm1}), (\ref{eq:eom1}) and (\ref{eq:eom2}) immediately yield the following EOM,
\be
	\left( \partial^2 + m_\Phi^2 -4 i g \, \sigma_2 U_2^\mu \partial_\mu \right) \Phi + \frac{1}{3!}\lambda \, \left( \Phi^T \Phi \Phi \right) = 0 \; ,
	\label{eq:eomsu2}
\ee
for constant fluid velocity and $\phi$. 
Comparing this result with Eq. (\ref{eq:nlkge}), contribution from the interacting fluid medium also appears in the third term but it contributes differently. Using Eqs. (\ref{eq:a}) and (\ref{eq:ae}) we arrive at non-relativistic limit,
\be
	\partial_\tau^2 \Phi + \partial_\x^2 \Phi - m_\Phi^2 \Phi + 4 g \sigma_2 \, \phi \left( \partial_\tau \Phi \mp i \, \v \cdot \partial_\x \Phi \right) - \frac{1}{3!}\lambda \, \left( \Phi \Phi^T \Phi \right) = 0 \; ,
	\label{eq:nreomsu2}
\ee
for $\v_2 = \v$. This is the nonlinear EOM governing the double stranded DNA dynamics with surrounding fluid medium in the present theory. The plus and minus signs show the dynamics of a strand and its counterpart surrounded by the fluids with opposite velocities.

Obviously, in contrast with the U(1) case it is hard to solve Eq. (\ref{eq:nreomsu2}) exactly. For the sake of simplification, let us consider 2-dimensional $(t,x)$ case of Eq. (\ref{eq:eomsu2}),
\be
  \frac{\pd^2 \Phi}{\pd t^2} - \frac{\pd^2 \Phi}{\pd x^2}
  + \alpha_t \frac{\pd \Phi}{\pd t} -  \alpha_x  \frac{\pd  \Phi}{\pd x}
  + m_\Phi^2 \Phi + \frac{\lambda}{3!} \Phi^3 =0 \; ,
  \label{eq:twostrain1}
\ee
where $\alpha_t \equiv -4 i g \sigma_2 \gamma \, \phi$ and $\alpha_x = 4 i g \sigma_2 \gamma \, v_x \, \phi$. Borrowing the traveling wave $\Phi(\xp) \equiv \Phi(x-Ct)$ as before we obtain,
\be
  \frac{d^2 \Phi}{d \xp^2} - \at \frac{d \Phi}{d \xp} + \mt_\Phi^2 \Phi + \lt \Phi^3 =0 \; ,
  \label{eq:twostrain4}
\ee
with $\at \equiv {(C \alpha_t + \alpha_x)}/{(C^2 - 1)}$, $\mt_\Phi^2 \equiv {m_\Phi^2}/{(C^2 - 1)}$ and  $\lt \equiv {\lambda}/{(3!(C^2 - 1))}$. For $\lt = 0$ it coincides with the equation of inharmonic oscillator, Eq. (\ref{eq:soliton}). 

For further simplification, we assume that $\lt$ is small enough such that the last term in Eq. (\ref{eq:twostrain4}) can be treated perturbatively. Then, we can expand the mass $\mt_\Phi$ in term of $\lt$, \ie $\mt_\Phi \simeq \mt_{\Phi_0} + \lt \mt_{\Phi_1}$, and $\Phi \simeq \Phi_0 + \lt \Phi_1$ up to $O(\lt)$ accuracy. Now we are ready to solve Eq. (\ref{eq:twostrain4}) order by order. 

The lowest order with respect to $\lt$, \ie $O(1)$, satisfies the following equation,
\be
 \frac{d^2 \Phi_0}{d \xp^2} - \at \frac{d \Phi_0}{d \xp} + \mt_{\Phi_0}^2 \Phi_0 = 0 \; .
  \label{eq:twostrain7}
\ee
Following the standard mathematical procedures the solution can in general be expressed in the form of $\Phi_0 = N_0^+ \mathrm{e}^{k_+ \xp} + N_0^- \mathrm{e}^{k_- \xp}$ with $k_\pm = \frac{1}{2} \left( \at \pm \sqrt{\at^2 - 4 \mt_{\Phi_0}^2}\right)$.
Therefore, the solution of Eq. (\ref{eq:twostrain7}) is simply,
\be
  \Phi_0 = {\displaystyle 2 N_0 \, \mathrm{e}^{\frac{\at}{2} \xp} \times} \left\{
  \begin{array}{lcl}
  1 & \mathrm{for} & \at^2 = 4 \mt_{\Phi_0}^2 \\
  {\displaystyle \cosh \left(\sqrt{\frac{\at^2}{4} - \mt_{\Phi_0}^2} \xp \right)} & \mathrm{for} & \at^2 > 4 \mt_{\Phi_0}^2 \\ 
  {\displaystyle \sin \left(\sqrt{\mt_{\Phi_0}^2 - \frac{\at^2}{4}} \xp \right)} & \mathrm{for} & \at^2 < 4 \mt_{\Phi_0}^2 \\
  \end{array}
  \right. \; ,
  \label{eq:solusi2}
\ee
by putting the normalization factor to be $N_0^+ = N_0^- \equiv N_0$. Each solution is corresponding to the over-damped, damped and regular harmonic oscillators respectively.

Subsequently, the next order, \ie $O(\lt)$, is governed by the equation,
\be
 	\frac{d^2 \Phi_1}{d \xp^2} - \at \frac{d \Phi_1}{d \xp} + \mt_{\Phi_0}^2 \Phi_1 
	+ 2 \mt_{\Phi_0} \mt_{\Phi_1} \Phi_0 + \Phi_0^3 = 0 \; .
  \label{eq:twostrain8}
\ee
The over-damped $\Phi_0$ in Eq. (\ref{eq:solusi2}) yields the general solution for $\Phi_1$ should be,
\be
\Phi_1 = N_1^+ \mathrm{e}^{\frac{\at}{2} \, \xp} + N_1^- \mathrm{e}^{\frac{3 \at}{2} \, \xp} \; .
\label{eq:solusis2od1}
\ee
Substituting the over-damped $\Phi_0$ and Eq. (\ref{eq:solusis2od1}) into Eq. (\ref{eq:twostrain8}), we obtain $N_1^- = - {\left( 32 N_0^3 \right)}/{\left( 3 \at^2 + 4 \mt_{\Phi_0}^2\right)}$. In the present case the first term in Eq. (\ref{eq:solusis2od1}) is vanishing for any $N_1^+$ since $\at^2 = 4 \mt_{\Phi_0}^2$. This also leads to the result $\mt_{\Phi_1} = 0$ since $\mt_{\Phi_0}, N_0 \neq 0$. Finally,
\be
	\Phi_1 = - \frac{32 N_0^3}{3 \at^2 + 4 \mt_{\Phi_0}^2} \mathrm{e}^{\frac{3 \at}{2}  \xp} \; ,
\ee
and $\mt_\Phi = \mt_{\Phi_0}$.

In the second case of damped $\Phi_0$, we make use of the equality $\cosh^3 x = 1/2 \cosh(3x) + \cosh x$ to obtain, 
\bea
  {\displaystyle \frac{d^2 \Phi_1}{d \xp^2} - \at \frac{d \Phi_1}{d \xp} + \mt_{\Phi_0}^2 \Phi_1 
  + 4 N_0^3 \mathrm{e}^{\frac{3 \at}{2} \xp} \, \cosh\left(3 \sqrt{\frac{\at^2}{4} - \mt_{\Phi_0}^2} \xp \right)} & &
  \nonumber \\
  + {\displaystyle 4 N_0^3 \mathrm{e}^{\frac{3 \at}{2} \xp} \left( 2 + 
 \frac{\mt_{\Phi_0}\mt_{\Phi_1}}{N_0^2} \mathrm{e}^{-\at \xp} \right) \cosh\left(\sqrt{\frac{\at^2}{4} - \mt_{\Phi_0}^2} \xp \right)} & = & 0 \; .
  \label{eq:solusis22}
\eea
Since $\cosh x < \cosh(3x)$ and the last term is enhanced only by a factor of as small as 2, Eq. (\ref{eq:solusis22}) can be approximately reduced to be, 
\be
  {\displaystyle \frac{d^2 \Phi_1}{d \xp^2} - \at \frac{d \Phi_1}{d \xp} + \mt_{\Phi_0}^2 \Phi_1 
  + 4 N_0^3 \mathrm{e}^{\frac{3 \at}{2} \xp} \, \cosh\left(3 \sqrt{\frac{\at^2}{4} - \mt_{\Phi_0}^2} \xp \right)} \simeq 0 \; .
  \label{eq:solusis23}
\ee
The solution is given by,
\be 
  \Phi_1 = \mathrm{e}^{\frac{3 \at}{2} \xp} \left[ N_1^+ \cosh \left(3 \sqrt{\frac{\at^2}{4} - \mt_{\Phi_0}^2} \xp \right)
   + N_1^- \sinh \left(3 \sqrt{\frac{\at^2}{4} - \mt_{\Phi_0}^2} \xp \right)\right] \; .
  \label{eq:solusis24}
\ee
Again substituting it into Eq. (\ref{eq:solusis23}) yields,
\bea
N_1^+ & = & -\frac{9}{4} N_0^3 \frac{\at^2 - 3 \mt_{\Phi_0}^2}{3 \at^4 + 36 \mt_{\Phi_0}^4 - 20 \at^2 \mt_{\Phi_0}^2} \; , \\
N_1^- & = & -\frac{9}{8} N_0^3 \frac{\at \sqrt{\at^2 - 4 \mt_{\Phi_0}^2}}{3 \at^4 + 36 \mt_{\Phi_0}^4 - 20 \at^2 \mt_{\Phi_0}^2} \; .
\eea
From these results, for $\xp > 0$ we can safely omit the sub-dominant $\sinh$ term in Eq. (\ref{eq:solusis24}), also because $N_1^+ > 2 N_1^-$ since $\at^2 > 4 \mt_{\Phi_0}^2$. Hence, 
\be 
  \Phi_1 = -\frac{9}{4} N_0^3 \frac{\at^2 - 3 \mt_{\Phi_0}^2}{3 \at^4 + 36 \mt_{\Phi_0}^4 - 20 \at^2 \mt_{\Phi_0}^2} \mathrm{e}^{\frac{3 \at}{2} \xp}  \cosh \left(3 \sqrt{\frac{\at^2}{4} - \mt_{\Phi_0}^2} \xp \right) \; .
  \label{eq:solusis25}
\ee

\begin{figure}[t]
        \centering 
	\includegraphics[width=11cm,angle=0,trim=0 0 0 0]{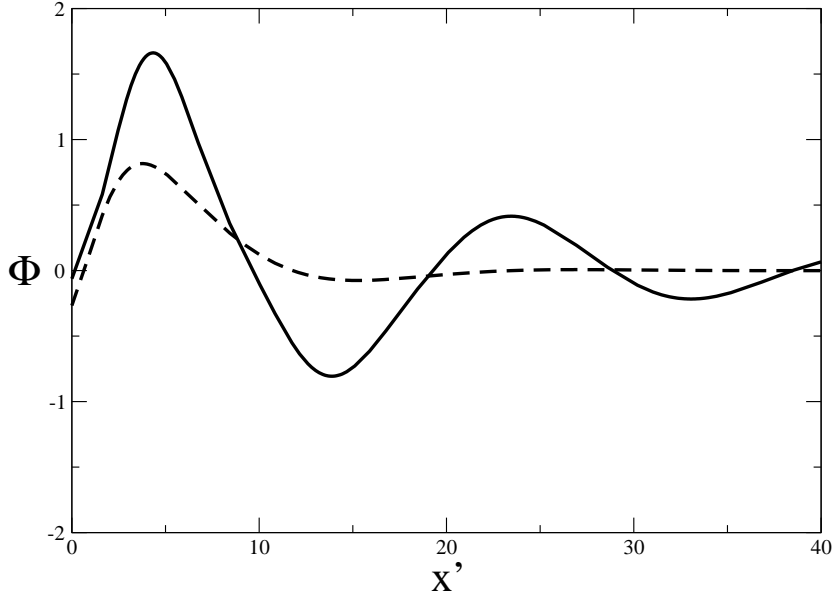}
        \caption{The wave function for 2-dimensional double stranded DNA as functions of $x^\prime$ with $v_x = 1.9$ (solid line) and $v_x = 1.7$ (dashed line) for a fixed parameter sets $(m_{\Phi_0},\phi,C,\lambda,g,N_0) = (1,1,2,4,1,1)$.}
        \label{fig:su2}
\end{figure}

The last case of regular harmonic oscillator is governed by the following equation,
\bea
  {\displaystyle \frac{d^2 \Phi_1}{d \xp^2} - \at \frac{d \Phi_1}{d \xp} + \mt_{\Phi_0}^2 \Phi_1 
  - 2 N_0^3 \mathrm{e}^{\frac{3 \at}{2} \xp} \, \sin\left(3 \sqrt{\mt_{\Phi_0}^2 - \frac{\at^2}{4}} \xp \right)} & &
  \nonumber \\
  + {\displaystyle 2 N_0^3 \mathrm{e}^{\frac{3 \at}{2} \xp} \left( 3 + 
 \frac{\mt_{\Phi_0}\mt_{\Phi_1}}{N_0^2} \mathrm{e}^{-\at \xp} \right) \sin\left(\sqrt{\mt_{\Phi_0}^2 - \frac{\at^2}{4}} \xp \right)} & = & 0 \; ,
  \label{eq:solusiho1}
\eea
using the relation $4 \sin^3 x = 3 \sin x - \sin (3x)$. In contrary with previous cases the general solution for Eq. (\ref{eq:solusiho1}) is complicated. So, let us assume here that the more rapid oscillation term, \ie the fourth term,  is dominant than the last one which reduces the equation to be, 
\be
  {\displaystyle \frac{d^2 \Phi_1}{d \xp^2} - \at \frac{d \Phi_1}{d \xp} + \mt_{\Phi_0}^2 \Phi_1 
  - 2 N_0^3 \mathrm{e}^{\frac{3 \at}{2} \xp} \, \sin\left(3 \sqrt{\mt_{\Phi_0}^2 - \frac{\at^2}{4}} \xp \right)} = 0 \; ,
  \label{eq:solusiho2}
\ee
Hence the general solution is simply,
\be 
  \Phi_1 = \mathrm{e}^{\frac{3 \at}{2} \xp} \left[ N_1^+ \cos \left(3 \sqrt{\mt_{\Phi_0}^2 - \frac{\at^2}{4}} \xp \right)
   + N_1^- \sin \left(3 \sqrt{\mt_{\Phi_0}^2 - \frac{\at^2}{4}} \xp \right)\right] \; .
  \label{eq:solusiho3}
\ee
Following similar procedures as before, 
\bea
N_1^+ & = & \frac{4}{3} N_0^3 \frac{\at \sqrt{4 \mt_{\Phi_0}^2 - \at^2}}{3 \at^4 + 36 \mt_{\Phi_0}^2 - 16 \at^2 \mt_{\Phi_0}^2} \; , \\
N_1^- & = & \frac{8}{3} N_0^3 \frac{\at^2 - 3 \mt_{\Phi_0}^2}{3 \at^4 + 36 \mt_{\Phi_0}^2 - 16 \at^2 \mt_{\Phi_0}^2} \; .
\eea
In non-relativistic case, by definition the condition $m_{\Phi_0} > \left| 2 g (C - v_x) \phi \right|$ should be fulfilled. Obviously, for large enough $\mt_{\Phi_0}$ ($\at^2 \ll 4 \mt_{\Phi_0}^2$ or $m_{\Phi_0} \gg \left| 2 g (C - v_x) \phi \right|$) the solution is dominated by $N_1^-$ term, while both terms are comparable for $\at^2 \rightarrow 4 \mt_{\Phi_0}^2$ or $m_{\Phi_0} \rightarrow \left| 2 g (C - v_x) \phi \right|$. These arguments lead to the result,
\be 
  \Phi_1 = \frac{4}{3} N_0^3 \frac{\at \sqrt{4 \mt_{\Phi_0}^2 - \at^2}}{3 \at^4 + 36 \mt_{\Phi_0}^2 - 16 \at^2 \mt_{\Phi_0}^2} \mathrm{e}^{\frac{3 \at}{2} \xp} \cos \left(3 \sqrt{\mt_{\Phi_0}^2 - \frac{\at^2}{4}} \xp \right) \; .
  \label{eq:solusiho4}
\ee

We should remark that up to the current accuracy there is no need in all cases to calculate the leading order of mass, $\mt_{\Phi_1}$. As a typical example, the wave function for the harmonically oscillating, \ie the sum of Eqs. (\ref{eq:solusi2}) and (\ref{eq:solusiho4}), double stranded DNA is depicted in Fig. \ref{fig:su2} as a function of $x^\prime$ for certain velocities. It can also be seen that the oscillation is sensitive to the fluid velocity.

\section{Summary and discussion}

We have introduced a new type of model to describe DNA using the gauge invariant fluid lagrangian. The lagrangian is able to accomodate various internal motions of DNA, from the single bulk to the double stranded of DNA as done in the preceeding section. The EOM's and its solutions for two typical cases using the Abelian U($1$) and non-Abelian SU($2$) lagrangians have been derived and investigated.

In the case of Abelian U($1$) lagrangian, we have seen from Eq. (\ref{eq:soliton}) that the interacting fluid medium characterized by the coupling constant $g$ influences the magnitude and also the width (associated to the dispersion or steppening rate) of solitonic wave equation as well, but regardless with the fluid velocity. On the other hand, obviously the matter self-interaction represented by its coupling constant $\lambda$ could change only the magnitude and not the dispersion or steppening rate of soliton. Actually, this provides a natural explanation for small and large amplitude regions of the internal motion of a single bulk of DNA immersed in bio-fluid without adding any new terms by hand as done in some previous works \cite{cwlim}. Furthermore, that contribution shifts the matter mass $m_\Phi$ to be $\mb_\Phi$. This is the so-called running mass induced by the dynamical fluctuation of internal kinematics in the system as a result of interaction between matter and fluid. However, the result is again independent on the fluid velocity.

In the second case, using the non-Abelian SU($2$) lagrangian we have constructed a model for double stranded DNA  in detail up to the level of its constituents, except for sugar and base composing the nucleoside. It has been shown that the EOM follows a similar form as in the U($1$) model, but the interacting fluid medium contributes in different way. The model requires that the DNA polymer would exist if and only if it resides in a fluid medium, represented by $I_2$ and $I_3$ bounds realized by fluid-matter interactions. Otherwise, the binding interactions $I_2$ and $I_3$ would vanish and the strands are broken. These results could be used to explain the deformation of DNA molecules associated with vanishing interactions in $I_{1,2,3}$. In contrast with the previous case, the fluid velocity plays an important role and changes the dynamics drastically, namely the highly damped, damped and regular harmonic oscillators. This supports a conclusion obtained in \cite{aslin}, that is the effect of hydrodynamic interactions on the dynamics of DNA translocation depends on the fluid velocity.

As mentioned earlier, both strands in a double stranded molecule are considered to follow the same EOM as Eq. (\ref{eq:nreomsu2}) with opposite sign of fluid velocities. In contrary, the single fragments of DNA belong to those strands are governed by Eq. (\ref{eq:nlkge}) independent on the fluid velocity, and  should behave identically no matter with the directions of its surrounding fluid velocities.

Further studies can be done using the lattice gauge simulation to calculate numerically, for instance the finite temperature partition function density $\z = \mathrm{exp}(1/T \int \mathrm{d}^3x \l)$. This kind of macroscopic ensemble provides direct relation between the internal dynamics of DNA and some physical observables like temperature and so on. Actually this is the main advantage of deploying the gauge invariant lagrangian like the present one. Such numerical calculations would be able to simulate quantitatively some phenomena in DNA like critical temperature or pressure related to the deformation of DNA molecules, etc. For example, one can investigate the critical temperature as a double stranded DNA is splitted into single strands \cite{gerland}, \ie $I_1 \rightarrow 0$ in the present model. Such studies are in the progress.

\section*{Acknowledgment}

We greatly appreciate fruitful discussion with T.P. Djun throughout the work. AS
thanks the Group for Theoretical and Computational Physics LIPI for warm
hospitality during the work. This work is
partially funded by the Indonesia Ministry of Research and Technology and the
Riset Kompetitif LIPI in fiscal years 2009 and 2010 (Contract no. 
11.04/SK/KPPI/II/2009 and 11.04/SK/KPPI/II/2010).

\end{document}